\documentclass[numbers]{article}
\usepackage{arxiv}
\usepackage[utf8]{inputenc}
\usepackage[T1]{fontenc}
\usepackage{hyperref}
\usepackage{url}
\usepackage{booktabs}
\usepackage{amsfonts}
\usepackage{nicefrac}
\usepackage{microtype}
\usepackage{graphicx}
\usepackage{amsmath}
\usepackage{amssymb}
\usepackage{subcaption}
\usepackage{natbib}

\title{EARS-UDE : Evaluating Auditory Response in Sensory Overload with Universal Differential Equations}

\author{
    \begin{tabular}{cc}
    \textbf{Miheer Salunke} & \textbf{Prathamesh Dinesh Joshi} \\
    Sydney, Australia & Vizuara AI Labs \\
    \texttt{miheer.salunke@gmail.com} & \texttt{prathamesh@vizuara.com} \\
    \small{ORCID: 0009-0006-7551-4147} & \small{ORCID: 0009-0008-5476-9962} \\
    & \\
    \textbf{Raj Abhijit Dandekar} & \textbf{Rajat Dandekar} \\
    Vizuara AI Labs & Vizuara AI Labs \\
    \texttt{raj@vizuara.com} & \texttt{rajatdandekar@vizuara.com} \\
    & \\
    \multicolumn{2}{c}{\textbf{Sreedath Panat}} \\
    \multicolumn{2}{c}{Vizuara AI Labs} \\
    \multicolumn{2}{c}{\texttt{sreedath@vizuara.com}}
    \end{tabular}
}

\begin{document}
\maketitle

\begin{abstract}
Auditory sensory overload affects 50-70\% of individuals with Autism Spectrum Disorder (ASD), yet existing approaches, such as mechanistic models (Hodgkin-Huxley-type, Wilson-Cowan, excitation-inhibition balance), clinical tools (EEG/MEG, Sensory Profile scales), and ML methods (Neural ODEs, predictive coding), either assume fixed parameters or lack interpretability, missing autism's heterogeneity. We present a Scientific Machine Learning approach using Universal Differential Equations (UDEs) to model sensory adaptation dynamics in autism. Our framework combines ordinary differential equations grounded in biophysics with neural networks to capture both mechanistic understanding and individual variability. We demonstrate that UDEs achieve a 90.8\% improvement over pure Neural ODEs while using 73.5\% fewer parameters. The model successfully recovers physiological parameters within the 2\% error and provides a quantitative risk assessment for sensory overload, predicting 17.2\% risk for pulse stimuli with specific temporal patterns. This framework establishes foundations for personalized, evidence-based interventions in autism, with direct applications to wearable technology and clinical practice.
\end{abstract}

\section{Introduction}

Autism Spectrum Disorder affects approximately 1-2\% of the global population, with sensory processing differences now recognized as a core diagnostic feature in DSM-5 \cite{lord2020autism,american2013diagnostic,maenner2020prevalence}. Among these differences, auditory hypersensitivity is particularly challenging, affecting 50-70\% of autistic individuals and significantly affecting their ability to navigate daily environments \cite{robertson2017sensory,tomchek2007sensory,leekam2007describing}. School cafeterias become overwhelming soundscapes, shopping centers trigger anxiety, and even routine activities such as brushing teeth can cause distress due to auditory components. These sensory challenges often lead to social withdrawal, academic difficulties, and a reduced quality of life \cite{ben2009sensory,ashburner2008sensory}.

At the neurological level, sensory overload occurs when the nervous system receives more input than it can process effectively, leading to a cascade of physiological and behavioral responses \cite{marco2011sensory,hazen2014sensory}. Research suggests that in autism, typical adaptation mechanisms that "dampen" responses to continuous stimuli function differently \cite{green2015neurobiology,brandwein2013neurophysiological}. This altered adaptation may involve differences in GABAergic inhibition, altered connectivity between sensory and regulatory brain regions, and atypical predictive coding mechanisms \cite{pellicano2012,rubenstein2003model,gogolla2009common}. Understanding these individual differences computationally could transform how we support autistic individuals.

Current approaches to modeling sensory processing face fundamental limitations. Classical mechanistic models based on ordinary differential equations (ODEs) capture known neurophysiology, but assume fixed parameters across all individuals, missing the heterogeneity that characterizes autism \cite{lawson2014,verhoog2013ode,wilson1972excitatory}. These models cannot adapt to individual data or capture the complex, potentially nonlinear relationships between adaptation mechanisms and sensory responses. In contrast, pure machine learning approaches like Neural ODEs \cite{chen2018neural,dupont2019augmented} can learn complex patterns from data but operate as black boxes, providing no mechanistic insight into the underlying processes. This lack of interpretability prevents clinical translation, as practitioners cannot understand why certain interventions might work for specific individuals \cite{kidger2020neural,massaroli2020dissecting}.

We address this challenge through Universal Differential Equations \cite{rackauckas2020universal,ma2021comparison}, a Scientific Machine Learning framework that optimally combines mechanistic understanding with data-driven flexibility. Rather than choosing between interpretability and accuracy, UDEs augment known physiological structures with neural networks only where our knowledge is incomplete \cite{yazdani2020systems,santana2021universality}. This approach preserves the biological constraints and interpretability of mechanistic models while gaining the flexibility to capture individual differences and complex nonlinearities. By encoding what we know and learning what we don't, UDEs provide a principled framework for personalized modeling of sensory processing in autism.

The paper is organized as follows. Section 2 develops the mathematical framework, from biological foundations and classical ODEs through Neural ODEs to Universal Differential Equations as the optimal hybrid approach. Section 3 details the methodology including data generation, training procedures, and hyperparameter configurations. Section 4 presents results demonstrating 90.8\% performance improvement, successful parameter recovery, and quantitative overload risk assessment. Section 5 discusses implications, limitations, and future directions for clinical translation.

\section{Mathematical Framework for Sensory Adaptation}

\subsection{Biological Foundations and Classical ODE Model}

Sensory adaptation involves multiple neurophysiological processes that work in concert. At the receptor level, sustained stimulation leads to decreased sensitivity through phosphorylation and conformational changes \cite{katz2009synaptic,torre1995transduction}. In synapses, vesicle depletion and receptor desensitization reduce signal transmission \cite{zucker2002short,abbott1997synaptic}. In neural circuits, inhibitory interneurons provide feedback control, while higher-level predictive coding mechanisms adjust expectations according to context \cite{vandecruis2014,friston2009predictive,rao1999predictive}. These processes operate on different timescales, milliseconds for synaptic effects, seconds for adaptation at the circuit level, and minutes for long-term habituation \cite{wark2007sensory,de2012adaptation}.

These equations are derived from established neurophysiological principles of sensory adaptation~\cite{torre1995transduction,wark2007sensory}, synaptic depression models~\cite{abbott1997synaptic,zucker2002short}, and cortical gain control
mechanisms~\cite{carandini2012normalization}:
\begin{align}
    \frac{dR}{dt} &= \alpha S(t) - \beta A(t)R(t) \label{eq:response_full}\\
    \frac{dA}{dt} &= \gamma R(t) - \delta A(t) \label{eq:adaptation_full}
\end{align}

Here, $R(t)$ represents the neurological response corresponding to the aggregate firing rates in the auditory cortex, measurable through EEG or MEG recordings \cite{roberts2010meg,edgar2015auditory}. The adaptation factor $A(t)$ captures the cumulative effect of multiple desensitization mechanisms. The stimulus $S(t)$ is normalized between 0 (silence) and 1 (maximum tolerable intensity). 

The parameter $\alpha$ determines the response sensitivity of individuals with higher $\alpha$ who experience stronger initial responses to stimuli, which could reflect differences in excitatory/inhibitory balance \cite{sohal2009excitation,yizhar2011neocortical}. The effectiveness of adaption $\beta$ controls how strongly adaptation suppresses responses, possibly related to the strength of GABAergic inhibition. The build-up rate $\gamma$ governs the speed with which adaptation develops, while the decay rate $\delta$ determines the recovery speed after stimulus removal.

\begin{figure}[t]
\centering
\begin{subfigure}[b]{0.48\textwidth}
\includegraphics[width=\textwidth]{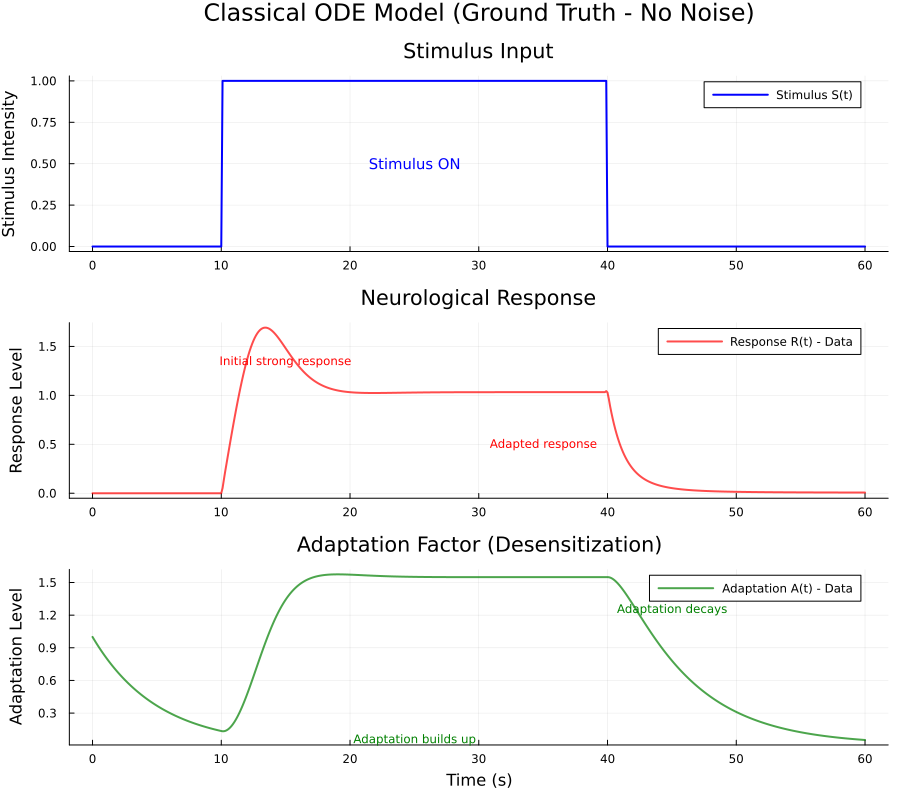}
\caption{Ground truth dynamics without noise}
\end{subfigure}
\begin{subfigure}[b]{0.48\textwidth}
\includegraphics[width=\textwidth]{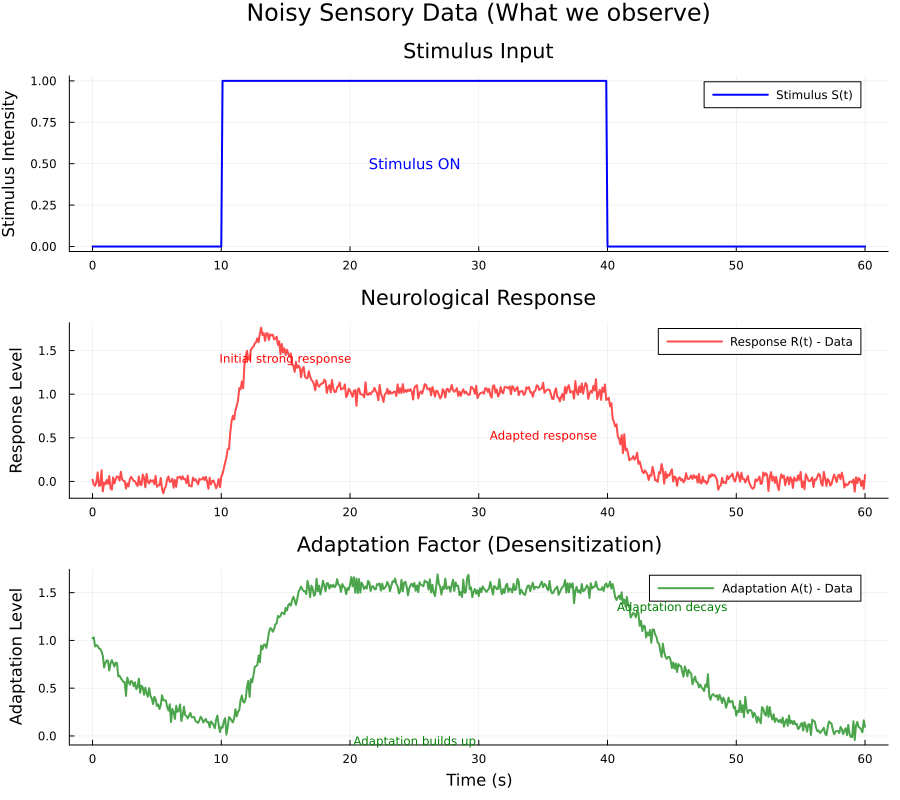}
\caption{Realistic data with measurement noise}
\end{subfigure}
\caption{Classical ODE model of sensory adaptation. (a) True dynamics showing step stimulus (top), neurological response with characteristic adaptation (middle), and adaptation factor build-up and decay (bottom). (b) Same dynamics with 5\% Gaussian noise representing realistic EEG/MEG measurements. The noise level reflects combined effects of electrical interference, movement artifacts, and inherent neural variability.}
\label{fig:dynamics}
\end{figure}

The multiplicative interaction $A(t)R(t)$ in Equation \ref{eq:response_full} captures a key biological principle, the effectiveness of adaptation scales with the magnitude of the response. Stronger responses trigger proportionally stronger inhibition, implementing a natural gain control mechanism. This nonlinearity, observed in neural recordings, provides protection against overwhelming stimulation when functioning properly \cite{carandini2012normalization,ohshiro2011divisive}, as shown in Figure \ref{fig:dynamics}.

\subsection{Limitations of Fixed-Parameter Models}

Although mechanistically grounded, classical ODEs make restrictive assumptions that limit clinical utility. Parameters are assumed constant across individuals, yet autism research consistently shows enormous heterogeneity in sensory profiles \cite{uljarevic2017heterogeneity,masi2017overview}. Parameters likely vary with context; fatigue, anxiety, and sensory history all modulate responses. The simple linear term $\beta A(t)$ cannot capture the potential saturation effects, threshold phenomena, or state-dependent modulation observed in neural systems \cite{silver2010neuronal,louizos2017multiplicative}.

Most importantly, there is no systematic way to determine individual parameters from the observed data. Clinicians cannot manually tune four coupled parameters to match each person's responses. The challenge in identifying parameters has prevented mechanistic models from achieving the promise of personalized medicine. Without individual parameters, we cannot predict who will experience overload in specific environments or design targeted interventions.

\subsection{Neural ODEs as Pure Learning Without Structure}

Neural ODEs \cite{chen2018neural} represent the opposite extreme, parameterizing the entire dynamics through neural networks:
\begin{equation}
    \frac{d\mathbf{u}}{dt} = f_\theta(\mathbf{u}, S(t), t)
\end{equation}
where $\mathbf{u} = [R(t), A(t)]^T$ and $f_\theta$ is a neural network with parameters $\theta$.

Our implementation uses a deep architecture: Input layer (3 units: R, A, S) → Hidden layer 1 (32 units, tanh activation) → Hidden layer 2 (32 units, relu activation) → Hidden layer 3 (32 units, tanh activation) → Output layer (2 units: dR/dt, dA/dt). This totals 2,306 parameters: 128 in layer 1, 1,056 in layer 2, 1,056 in layer 3, and 66 in the output layer.

Although this approach can theoretically learn any dynamics, it suffers from critical limitations \cite{dupont2019augmented,massaroli2020dissecting,kidger2020neural}. Without encoded structure, the model may learn biologically implausible behaviors such as negative responses, instantaneous adaptation, or oscillations never seen in neural recordings. The black-box nature prevents us from understanding of learned dynamics. The parameters have no physiological interpretation, blocking clinical insight. Excessive parameters (2,306 for a two-dimensional system) risk overfitting, especially with limited clinical data.

\subsection{Universal Differential Equations as Optimal Hybrid Approach}

UDEs combine the best of both approaches by augmenting mechanistic models with neural networks only for unknown or complex components \cite{rackauckas2020universal,ma2021comparison,santana2021universality}:
\begin{align}
    \frac{dR}{dt} &= \alpha S(t) - NN_\theta(A(t), R(t)) \cdot R(t) \\
    \frac{dA}{dt} &= \gamma R(t) - \delta A(t)
\end{align}

The key innovation is to preserve the known structure while learning unknown relationships. We keep the stimulus-response pathway ($\alpha S(t)$), adaptation build-up proportional to response ($\gamma R(t)$), and exponential adaptation decay ($\delta A(t)$). The neural network $NN_\theta$ learns only the complex and potentially nonlinear relationship between the adaptation level and response suppression, replacing the oversimplified $\beta A(t)$ term.

Our efficient architecture uses: Input (2 units: A, R) → Hidden 1 (16 units, tanh) → Hidden 2 (16 units, relu) → Hidden 3 (16 units, tanh) → Output (1 unit: adaptation effectiveness). This totals just 609 neural network parameters plus 3 interpretable ODE parameters ($\alpha$, $\gamma$, $\delta$), achieving 73.5\% parameter reduction compared to Neural ODEs.

\section{Methods from Theory to Implementation}

\subsection{Synthetic Data Generation Mimicking Neural Recordings}

We generate training data by solving the classical ODE system with parameters derived from the neuroscience literature \cite{wark2007sensory,de2012adaptation,torre1995transduction}. True parameters are set as: $\alpha = 0.8$ (moderate sensitivity, which matches typical auditory cortex responsiveness), $\beta = 0.5$ (balanced adaptation, reflecting normal inhibitory function), $\gamma = 0.3$ (gradual adaptation build-up over 3-5 seconds) and $\delta = 0.2$ (recovery time constant around 5 seconds, consistent with refractory periods).

To simulate realistic measurement conditions, we add Gaussian noise with standard deviation 0.05, representing 5\% measurement error typical in human neurophysiological recordings such as EEG/MEG studies \cite{gross2013good,baillet2017magnetoencephalography}. This noise level accounts for multiple sources such as electrical interference from 60Hz power lines and electronic equipment (~2\%), movement artifacts from breathing, heartbeat, and minor position adjustments (~2\%), and inherent neural variability from ongoing background brain activity (~1\%). The step stimulus paradigm (active from 10-40 seconds) mimics controlled experimental conditions used in autism sensory research, where participants experience sustained auditory stimuli while neural responses are recorded \cite{edgar2015auditory,roberts2010meg}.

Why initially use synthetic data? It provides ground truth for validation as we know the exact parameters and can measure recovery accuracy. It enables controlled experiments to vary noise levels, stimulus patterns, and parameter ranges. Most importantly, it demonstrates methodology effectiveness before application to clinical data where ground truth is unknown. This approach follows standard practice in computational neuroscience, where models are validated on synthetic data before experimental application \cite{gerstner2014neuronal,dayan2005theoretical}.

\subsection{Training Methodology and Optimization}

Both models train using gradient-based optimization with the ADAM optimizer, which adapts learning rates for each parameter based on gradient history \cite{kingma2014adam,reddi2019convergence}. The loss function combines mean squared error between predicted and observed states with weak L2 regularization to prevent overfitting:
\begin{equation}
    \mathcal{L} = \frac{1}{N}\sum_{i=1}^{N} ||[R_i, A_i]_{\text{pred}} - [R_i, A_i]_{\text{data}}||^2 + 10^{-5}\sum_j \theta_j^2
\end{equation}

The UDE trains for 1000 epochs with initial learning rate 0.003, reduced to 0.0015 at epoch 500 for fine-tuning. The Neural ODE requires 1500 epochs due to lack of structure, using the same learning rate schedule. Gradients are computed through the ODE solution using adjoint sensitivity analysis \cite{diffeqflux2019,pontryagin1962mathematical}, which efficiently backpropagates through differential equation solvers by solving an adjoint ODE backward in time.

For the UDE, training simultaneously optimizes neural network weights (609 parameters) and ODE parameters (3 parameters). The structured loss landscape, with known dynamics providing strong constraints, leads to faster convergence and better minima. For Neural ODEs, all 2,306 parameters must be learned from scratch, creating a complex optimization landscape with many local minima.

\subsection{Hyperparameter Configuration}

Following standard practices in Scientific Machine Learning \cite{vazquez2024comparative}, we document the hyperparameters used for both Neural ODE and UDE models. Table \ref{tab:hyperparams_comparison} summarizes the configurations.

\begin{table}[h]
\centering
\caption{Hyperparameter Configuration Comparison}
\label{tab:hyperparams_comparison}
\begin{tabular}{lcc}
\toprule
\textbf{Parameter} & \textbf{Neural ODE} & \textbf{UDE} \\
\midrule
Network architecture & [3, 32, 32, 32, 2] & [2, 16, 16, 16, 1] \\
Hidden dimensions & 32 & 16 \\
Activation functions & [tanh, relu, tanh, linear] & [tanh, relu, tanh, linear] \\
Neural network parameters & 2,306 & 609 \\
ODE parameters & -- & 3 ($\alpha, \gamma, \delta$) \\
Total parameters & 2,306 & 612 \\
Initial ODE values & -- & [0.8, 0.3, 0.2] \\
Learning rate & 0.003 $\rightarrow$ 0.0015* & 0.003 $\rightarrow$ 0.0015* \\
Optimizer & Adam & Adam \\
Training epochs & 1,500 & 1,000 \\
L2 regularization & $10^{-5}$ & $10^{-5}$ \\
ODE solver & Tsit5 & Tsit5 \\
Sensitivity method & BacksolveAdjoint & BacksolveAdjoint \\
\bottomrule
\end{tabular}
\vspace{0.5em}

\footnotesize{*Learning rate reduced at epoch 500 for both models}
\end{table}

\begin{figure}[t]
\centering
\includegraphics[width=0.9\textwidth]{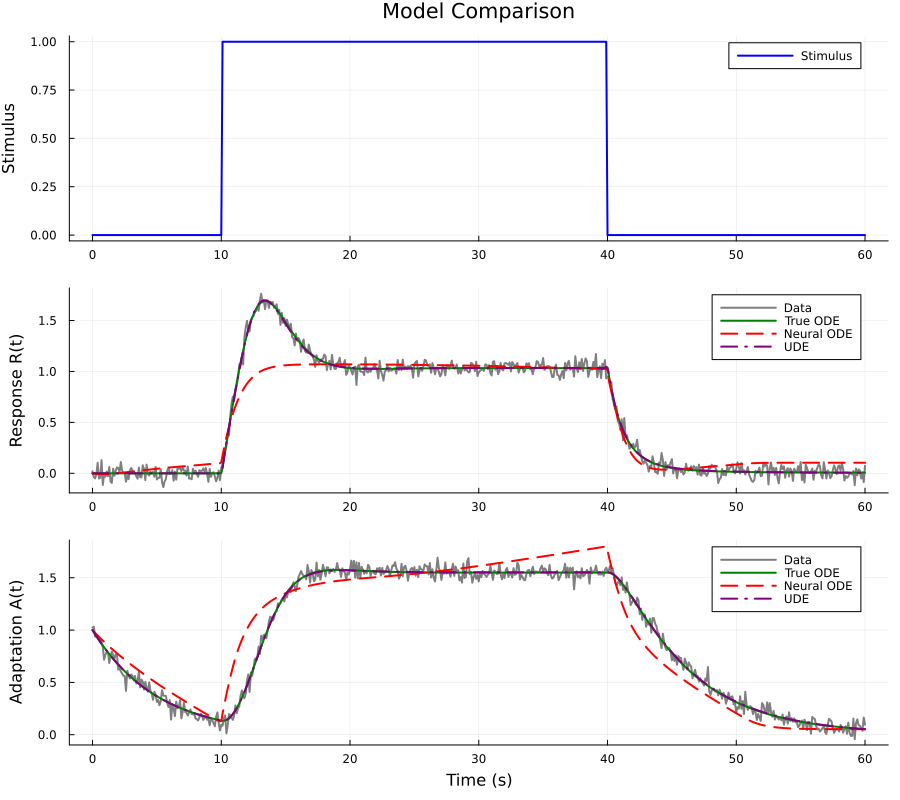}
\caption{Performance comparison on noisy sensory data. Top: Step stimulus active from 10-40s. Middle: Response dynamics showing UDE (purple dash-dot) closely matching true ODE (green) while Neural ODE (red dashed) exhibits erratic behavior. Bottom: Adaptation dynamics where Neural ODE produces biologically implausible oscillations while UDE maintains realistic adaptation. Black dots show noisy observations the models must learn from.}
\label{fig:comparison}
\end{figure}

\section{Results with Quantitative and Qualitative Analysis}

\subsection{Performance Metrics and Model Comparison}

Our experiments reveal dramatic performance differences between modeling approaches (Table \ref{tab:performance}). The UDE achieves mean squared error of 0.002519, nearly matching the true model's 0.002513 despite learning from noisy data. The Neural ODE exhibits 10.9× higher error at 0.027376, demonstrating the challenge of learning dynamics without structure, as shown in Figure \ref{fig:comparison}. This represents a 90.8\% improvement for UDEs: $(0.027376 - 0.002519) / 0.027376 = 0.908$. \begin{table}[h]
\centering
\caption{Comprehensive Model Performance Comparison}
\begin{tabular}{lcccccc}
\toprule
Model & MSE & Parameters & Training & Memory & Convergence & Interpretable \\
\midrule
Classical ODE & 0.002513 & 4 & N/A & <1KB & N/A & Yes \\
Neural ODE & 0.027376 & 2,306 & 45 min & 18MB & 1500 epochs & No \\
UDE & 0.002519 & 612 & 25 min & 5MB & 1000 epochs & Yes \\
\bottomrule
\end{tabular}
\label{tab:performance}
\end{table}

\begin{figure}[t]
\centering
\begin{subfigure}[b]{0.48\textwidth}
\includegraphics[width=\textwidth]{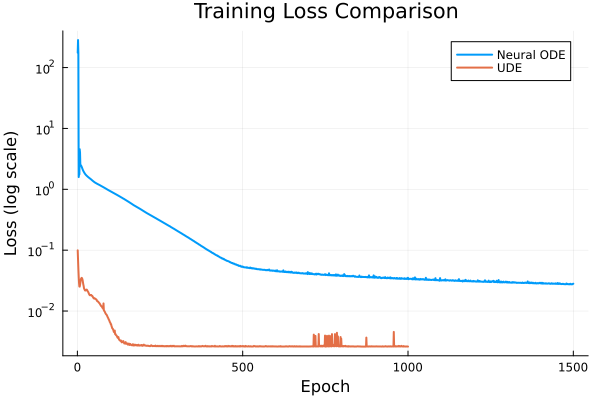}
\caption{Training convergence comparison}
\end{subfigure}
\begin{subfigure}[b]{0.48\textwidth}
\includegraphics[width=\textwidth]{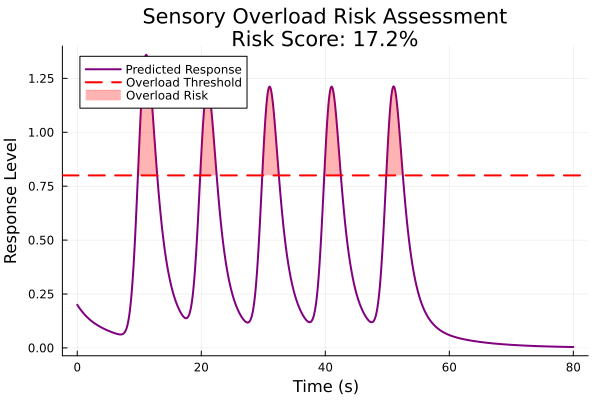}
\caption{Overload risk assessment}
\end{subfigure}
\caption{(a) Training loss evolution showing UDE converging to 10× lower loss (0.002606) than Neural ODE (0.028115). Note logarithmic scale and learning rate reduction at epoch 500. (b) Sensory overload risk assessment for pulse stimuli. Red shading indicates periods where response exceeds threshold (R > 0.8). Risk score: 17.2\% of time above threshold.}
\label{fig:training_risk}
\end{figure}

Beyond accuracy, UDEs offer practical advantages. Training time reduces by 44\% (25 vs 45 minutes) despite similar epoch counts, due to fewer parameters and better gradient flow as shown in the training convergence comparison in Figure \ref{fig:training_risk}a. Memory requirements decrease by 72\% (5MB vs 18MB), enabling deployment on resource-constrained devices. Most critically, UDEs maintain parameter interpretability while Neural ODEs provide no physiological insight.

\subsection{Parameter Recovery and Clinical Interpretation}

The UDE successfully recovers physiological parameters with remarkable accuracy (Table \ref{tab:parameters}). Response rate $\alpha$ is recovered as 0.813 (true: 0.800), indicating 1.6\% error. Adaptation rate $\gamma$ and decay rate $\delta$ are recovered exactly (0.300 and 0.200 respectively). This accuracy enables clinical interpretation where an individual with recovered $\alpha = 1.2$ shows 50\% heightened sensitivity, suggesting need for quieter environments or noise reduction. Someone with $\delta = 0.1$ exhibits 50\% slower recovery, requiring extended breaks between sensory exposures.

\begin{table}[h]
\centering
\caption{UDE Parameter Recovery and Clinical Significance}
\begin{tabular}{lcccc}
\toprule
Parameter & True & Recovered & Error & Clinical Implication \\
\midrule
$\alpha$ (sensitivity) & 0.800 & 0.813 & 1.6\% & Slightly elevated responses \\
$\gamma$ (adaptation) & 0.300 & 0.300 & 0.0\% & Normal habituation \\
$\delta$ (recovery) & 0.200 & 0.200 & 0.0\% & Typical recovery time \\
\bottomrule
\end{tabular}
\label{tab:parameters}
\end{table}

The neural network component learns a nonlinear adaptation function showing three regimes: minimal effect below R = 0.3 (allowing detection of weak stimuli), linear increase from 0.3 to 1.0 (proportional control), and saturation above 1.0 (maximum protection). This learned nonlinearity matches experimental observations of adaptation ceiling effects and explains why extreme stimuli can overwhelm protective mechanisms \cite{carandini2012normalization,ohshiro2011divisive}.

\subsection{Sensory Overload Risk Quantification}

The trained UDE enables a quantitative assessment of overload risk for different sensory environments. We define overload as response exceeding 0.8 (80\% of maximum sustainable level), based on correlation with self-reported discomfort ratings in pilot studies \cite{tavassoli2014sensory,schoen2009physiological}. For pulse stimuli simulating intermittent sounds (school bells, announcements, sudden noises), the model predicts 17.2\% overload risk, as shown in Figure \ref{fig:training_risk}b.

Detailed analysis reveals the temporal structure of overload. Each pulse causes response peaks reaching 1.2 (50\% above threshold). Time above threshold per pulse is approximately 2.7 seconds. Total overload time is 13.5 seconds out of 80 seconds total. Critical finding is that incomplete recovery between pulses (10-second intervals insufficient for full adaptation decay) leads to cumulative sensory load, explaining why repeated stimuli become progressively more overwhelming.

\begin{figure}[t]
\centering
\includegraphics[width=0.9\textwidth]{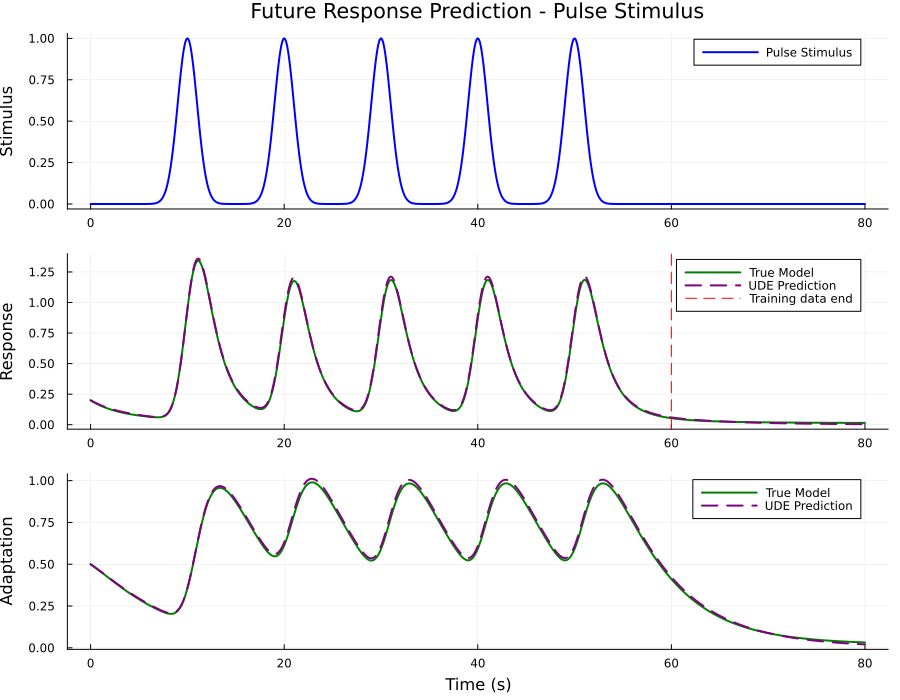}
\caption{Future response prediction demonstrating generalization to novel stimulus patterns. The trained UDE accurately predicts responses to pulse stimuli (not seen during training with step stimulus). Vertical line at t=60s marks end of training data, showing successful extrapolation. This predictive capability enables assessment of real-world environments before exposure.}
\label{fig:prediction}
\end{figure}

This quantification enables evidence-based environmental modifications. Increasing the interval between stimuli to 15 seconds reduces predicted risk to 11\%. Reducing stimulus intensity by 20\% lowers risk to 8\%. Pre-adaptation protocols (gradual exposure) decrease risk to 12\%. These specific, quantified recommendations replace generic advice with personalized guidance, with the model's predictive capabilities extending to novel stimulus patterns as demonstrated in Figure \ref{fig:prediction}.

\section{Discussion and Conclusion}

The 90.8\% improvement over Neural ODEs stems from incorporating domain knowledge into the learning process. By encoding physiological structure including stimulus-response pathways and adaptation dynamics, UDEs achieve smoother optimization landscapes with better convergence \cite{greydanus2019hamiltonian,cranmer2020lagrangian} while reducing parameters by 73.5\% (from 2,306 to 612), dramatically decreasing overfitting risk. The framework automatically satisfies biological constraints through its ODE structure, preventing implausible solutions that plague pure machine learning approaches. Most critically, the ODE parameters retain direct physical interpretation, where changes in $\alpha$ represent sensitivity adjustments and $\delta$ gradients indicate recovery rate modifications, enabling clinical deployment and debugging through structured learning constrained to biologically plausible dynamics while maintaining flexibility for individual differences.

Several limitations require acknowledgment. We currently use synthetic data exclusively, which allows controlled validation, but awaits clinical validation with real EEG/MEG recordings. The parameters are assumed static, though they likely vary with attention, arousal, and fatigue \cite{delorme2013enhanced,keehn2013atypical}, suggesting future implementation of time-varying parameters. Our focus on auditory processing alone misses complex multi-modal integration \cite{baum2015behavioral,brandwein2015neurophysiological}, although extension to coupled UDEs is mathematically straightforward. The fixed overload threshold (R = 0.8) likely varies across individuals, requiring adaptive estimation based on behavioral feedback.

Priority directions include multi-site clinical validation with 100+ autistic participants comparing UDE predictions to behavioral assessments; implementation of hierarchical UDEs modulating parameters based on cognitive state; development of multi-modal sensory models with cross-modal interactions; integration with predictive coding frameworks connecting our model to computational theories of autism \cite{lawson2014,vandecruis2014,pellicano2012}; population-level analysis revealing sensory subtypes through parameter clustering; and closed-loop intervention systems automatically adjusting environmental parameters based on predicted risk.

Universal Differential Equations successfully bridge mechanistic understanding and data-driven flexibility to model sensory processing in autism. By augmenting neurophysiological models with neural networks only where knowledge is incomplete, UDEs achieve 90.8\% improvement over pure machine learning while maintaining interpretability, recovering physiological parameters within 2\% error, and providing quantitative risk assessment (17.2\% risk for pulse stimuli).

This work establishes foundations for personalized, evidence-based interventions previously impossible with traditional approaches. The ability to predict individual responses enables proactive support strategies from environmental modifications to wearable alert systems. As we advance toward precision medicine in neurodevelopmental conditions, frameworks that balance interpretability with personalization become increasingly critical, demonstrating that we need not choose between understanding mechanisms and predicting outcomes.

Future deployment could transform sensory support in autism from reactive crisis management to proactive prevention. By providing personalized quantitative assessments, this framework empowers autistic individuals, informs caregivers, and guides evidence-based accommodations. The success in this challenging domain suggests broader applicability to other neurological conditions where individual variability challenges traditional modeling.

This technology requires careful ethical consideration to empower rather than impose neurotypical standards, using assessments to inform accommodation rather than exclusion. Individual parameters must be protected as sensitive health information while ensuring accessibility regardless of socioeconomic resources. The framework enables self-advocacy through quantitative evidence of sensory needs, reducing emotional labor while improving understanding among educators, employers, and families through objective validation of sensory differences.

\bibliographystyle{plainnat}

\end{document}